\documentclass[reprint,amsmath,amssymb,aip,showpacs,pra]{revtex4-1}
\usepackage{graphicx}
\usepackage{dcolumn}
\usepackage{bm}
\usepackage[usenames,dvipsnames]{color}
\usepackage[utf8]{inputenc}
\usepackage{hyperref}
\graphicspath{{./Files/PDFs/}}

\begin{document}


\title{Amplitude Mediated Chimera States with Active and Inactive Oscillators}

\author{Rupak Mukherjee} 
\email{rupakmukherjee01@gmail.com}
\author {Abhijit Sen}
\email{abhijit@ipr.res.in}
\affiliation{Institute for Plasma Research, HBNI, Bhat, Gandhinagar - 382428, India}

\begin{abstract}
The emergence and nature  of amplitude mediated {\it chimera} states, spatio-temporal patterns of co-existing coherent and incoherent regions, are investigated for a globally coupled system of active and inactive Ginzburg-Landau oscillators. The existence domain of such states is found to shrink and shift in parametric space as the fraction of inactive oscillators is increased. The role of inactive oscillators is found to be two fold - they get activated to form a separate region of coherent oscillations and in addition decrease the common collective frequency of the coherent regions by their presence. The dynamical origin of these effects is delineated through a bifurcation analysis of a reduced model system that is based on a mean field approximation. Our results may have practical implications for the robustness of such states in biological or physical systems where age related deterioration in the functionality of components can occur. 
\end{abstract}

\maketitle

{\bf{The chimera state, a novel spatio-temporal pattern of co-existing coherent and incoherent regions, was originally discovered as a collective state of a simple model system of identical phase oscillators that are non-locally coupled to each other. Since this original discovery, chimera states have been shown to occur in a wide variety of systems and have been the subject of intense theoretical and experimental studies. They serve as a useful paradigm for describing similar collective phenomena observed in many natural systems such as the variety of metastable collective states of the cortical neurons in the brain or uni-hemispheric sleep in certain birds and mammals during which one half of the brain is synchronized while the other half is in a de-synchronized state. An important question to ask is what happens to such states when some of the components of the system lose their functionality and die, or, in the context of the oscillator model, when some of the oscillators cease to oscillate. In our present work, we examine this issue by investigating the dynamics of a system consisting of a mix of active (oscillating) and inactive (non-oscillating) Ginzburg-Landau oscillators that are globally coupled to each other. The chimera states of this system, known as Amplitude Mediated Chimeras (AMCs) are a more generalized form of the original phase oscillator chimeras and display both amplitude and phase variations. We find that the inactive oscillators influence the AMCs in several distinct ways. The coupling with the rest of the active oscillators revives their oscillatory properties and they become a part of the AMC as a separate coherent cluster thereby modulating the structure of the AMC. Their presence also reduces the overall frequency of the coherent group of oscillators. Finally, they shrink the existence region of the AMCs in the parametric space of the system. A remarkable finding is that the AMCs continue to exist even in the presence of a very large fraction (about 90\%) of inactive oscillators which suggests that they are very robust against aging effects. Our findings can be practically relevant for such states in biological or physical systems where aging can diminish the functional abilities of component parts.}}\\

\section{Introduction}
The {\it chimera} state, a novel collective phenomenon observed in coupled oscillator systems, that spontaneously emerges as a spatio-temporal pattern of co-existing synchronous and asynchronous groups of oscillators, has attracted a great deal of attention in recent years \cite{Motter:10,Panaggio:15}. First observed and identified by Kuramoto and Battogtokh\cite{Kuramoto:02} for a model system of identical phase oscillators that are non-locally coupled, such states have now been shown to occur in a wide variety of systems \cite{Panaggio:15} and under less restrictive conditions than previously thought of \cite{Sethia:13,Sethia:14}. Theoretical and numerical studies have established the existence of chimera states in neuronal models \cite{Santos:17}, in systems with non-identical oscillators \cite{Montbrio:04,Laing:09}, time delay coupled systems \cite{Sethia:08} and globally coupled systems that retain the amplitude dynamics of the oscillators \cite{Sethia:13,Sethia:14}.  Chimera states have also been observed experimentally in chemical \cite{Tinsley:12,Nkomo:13}, optical \cite{Hagerstrom:12}, mechanical \cite{Martens:13}, electronic \cite{Gambuzza:14} and electro-chemical \cite{Wickramasinghe:13} oscillator systems. Furthermore, chimera states have been associated with some natural phenomena such as unihemispheric sleep \cite{Abrams:08} in certain birds and mammals during which one half of the brain is synchronized while the other half is in a de-synchronized state \cite{Rattenborg:00}. In fact, the electrical activity of the brain resulting in the collective dynamics of the cortical neurons provides a rich canvas for the application of chimera states and is the subject of many present day studies \cite{Shanahan:10,Wildie:12,Majhi:16}. 
It is believed that a prominent feature of the dynamics of the brain is the generation of a multitude of meta-stable chimera states that it keeps switching between. Such a process, presumably, is at the heart of our ability to respond to different stimuli and  to much of our learning behavior. Chimera states are thus vital to the functioning of the brain and it is important therefore to investigate their existence conditions and robustness to changes in the system environment. An interesting question to ask is how the formation of chimera states can be influenced by the loss of functionality of some of the constituent components of the system. In the case of the brain it could be the damage suffered by some of the neuronal components due to aging or disease \cite{Naik:17}. For a model system of coupled oscillators this can be the loss of oscillatory behavior of some of the oscillators. In this paper we address this question by studying the existence and characteristics of the amplitude mediated chimera (AMC) states in an ensemble of globally coupled Complex Ginzburg-Landau (CGL) oscillators some of which are in a non-oscillatory (inactive) state. The coupled set of CGL oscillators display a much richer dynamics \cite{Nicolaou:2017} than the coupled phase oscillator systems on which many past studies on chimeras have been carried out \cite{Battogtokh:2000}. For a set of locally coupled CGL equations (corresponding to the continuum limit) Nicolaou {\it et al} \cite{Nicolaou:2017} have shown the existence of a chimera state consisting of a coherent domain of a frozen spiral structure and an incoherent domain of amplitude turbulence. The nature of the incoherent state in this case is more akin to turbulent patches observed in fluid systems and distinctly different from the incoherent behaviour obtained in non-locally coupled discrete systems \cite{Battogtokh:2000}. For globally coupled systems, chimera states with amplitude variations have been obtained for a set of Stuart-Landau oscillators by Schmidt and Krischer \cite{Schmidt:2015}. However the coupling they use is of a nonlinear nature and the chimera states emerge as a result of a clustering mechanism. The AMC states, on the other hand, do not require any nonlinear coupling and exist in a system of linearly coupled CGL equations.  The conditions governing the emergence of the AMCs are thus free of the topological and coupling constraints of the classical phase oscillator chimeras and may therefore have wider practical applications.\\

Past studies on the robustness of the collective states of a population of coupled oscillators that are a mix of oscillatory and non-oscillatory (inactive) elements have been restricted to the synchronous state \cite{Daido:2004,Bhumika:14,Zou:13,Zou:15}. Our work extends such an analysis to the emergent dynamics of the amplitude mediated chimera state. We find that the presence of inactive elements in the system can significantly impact the existence domain and the nature of the AMCs. 
The existence domain of the AMC states is found to shrink and shift in parametric space as the fraction of inactive oscillators is increased in the system. Under the influence of the coupling the inactive oscillators experience a revival and turn oscillatory to form another separate coherent region. They also decrease the collective frequency of the coherent regions. These results are established through extensive numerical simulations of the coupled system and by a systematic comparison with past results obtained in the absence of the inactive oscillators. To provide a deeper understanding of the numerical results and to trace the dynamical origins of these changes we also present a bifurcation analysis of a reduced model system comprised of two single driven oscillators with a common forcing term obtained from a mean field approximation. 

The paper is organized as follows. In the next section (section II) we describe the full set of model equations and summarize the past results on the AMC states obtained from these equations in the absence of any inactive oscillators. Section III details our numerical simulation results obtained for various fractions of inactive oscillators and discusses the consequent modifications in the existence regions and characteristics of the AMC states. In section IV we provide some analytic results from a reduced model system based on a mean field theory to explain the findings of section III. Section V gives a brief summary and some concluding remarks on our results.


\section{Model Equations}
We consider a system of globally coupled complex Ginzburg-Landau type oscillators governed by the following set of equations
\begin{eqnarray}
&& \dot{W}_j = \alpha_j W_j - (1+iC_2)|W_j|^2W_j \nonumber\\
&& \label{CCGLE} ~~~~~~~~~~~~~~~~~~~~~~ + K(1+iC_1)(\overline{W}-W_j)\\
&& \overline{W} = \frac{1}{N}\sum\limits_{i=1}^N W_i
\end{eqnarray}
where, $W_{j}$ is the complex amplitude of the $j^{th}$ oscillator, $\overline{W}$ is the mean field, overdot represents a differentiation w.r.t. time ($t$), $C_1, C_2, K$ are real constants, $N$ is the total number of oscillators and $\alpha_j$ is a parameter specifying the distance from the Hopf bifurcation point. In the absence of coupling, the $j^{th}$ oscillator exhibits a periodic oscillation if $\alpha_j \textgreater 0$ while if $\alpha_j \textless 0$ then, the $j^{th}$ oscillator settles down to the fixed point $W_j = 0$ and does not oscillate. We therefore divide up the system of oscillators into two subsets: the oscillators with $j \in \{1, \cdots ,Np\}$ form one population of oscillators with $\alpha_j < 0$ and represent inactive oscillators, while the oscillators with $j \in \{Np+1, \cdots ,N\}$ have $\alpha_j >0$ and represent active oscillators. Here $p$ ($0 \leq p \leq 1$) represents the fraction of inactive oscillators in the system. We assume that our system size is sufficiently large that the ratio $p$ can be treated as a continuous variable \cite{Daido:2004}. 
For simplicity we also assume, 
$\alpha_j = -b $  $\forall$ $j$ $\in$ $\{1,\cdots,Np\}$ and $\alpha_j =  a$ $\forall$ $j$ $\in$ $\{Np+1, \cdots , N\}$, where both $a$ and $b$ are positive constants. 
The set of equations (\ref{CCGLE}), in the absence of the inactive oscillators ($p=0$) has been extensively studied in the past \cite{Chabanol:97} and shown to possess a variety of collective states including synchronous states, splay states, single and multi-cluster states, chaotic states and more recently, also, amplitude mediated chimera states (AMCs) \cite{Sethia:14}. The AMC states occur in a restricted region of the ($C_1-K-C_2$) parameter space and are often co-existent with stable synchronous states \cite{Sethia:14}. 
Our objective in this paper is to study the effect of inactive elements on the nature of the AMCs and of  modifications if any of their existence region. Accordingly, we carry out an extensive numerical exploration of the set of equations (\ref{CCGLE}) in the relevant parametric domain of ($C_1-K$) space with $C_2=2$ and compare them to previous results \cite{Sethia:14} obtained in the absence of inactive oscillators.


\section{Amplitude Mediated Chimera States}
\begin{figure}[h!]
\begin{center}
\includegraphics[scale=0.65]{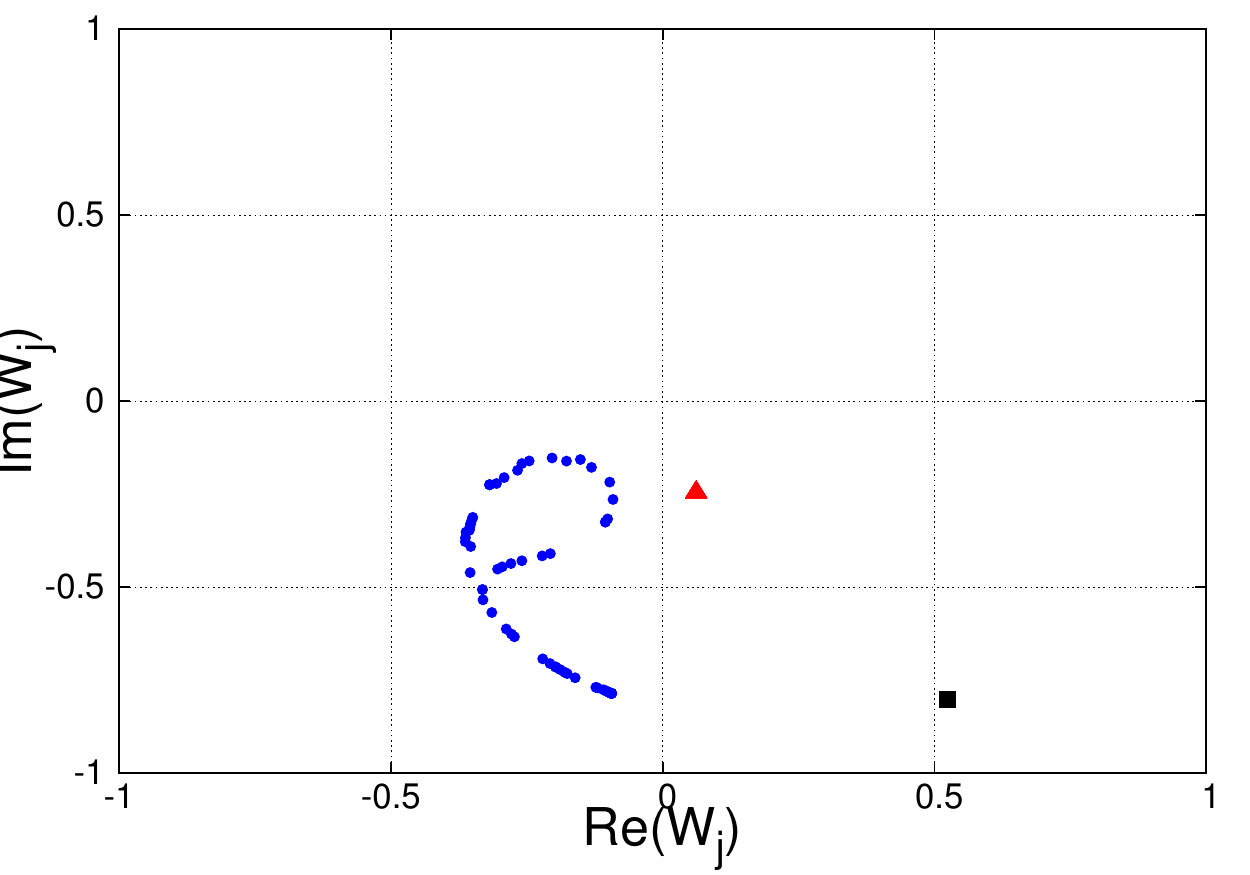}
\caption{(Color online) $N = 201, K = 0.7, C_1 = -1, C_2 = 2$, $a = b = 1$, $p = 0.1$. Blue points represent the incoherent active oscillators, the black square represents the coherent active oscillators and the red triangle represents the dead oscillators.}
\label{chimera_fig}
\end{center}
\end{figure}

In Fig.~(\ref{chimera_fig}) we display a snapshot of the time evolution of a typical AMC state obtained by a numerical solution of Eq.(\ref{CCGLE}) for $N=201, K = 0.7, C_1 = -1, C_2 = 2$, $a = b = 1$ and $p = 0.1$. 
We start with a state where the oscillators are uniformly distributed over a ring i.e.
\begin{eqnarray*}
Re(W_j)=\cos(2\pi j/N)\\
Im(W_j)=\sin(2\pi j/N)
\end{eqnarray*}
However the existence and formation of the AMC states are found to be independent of the initial conditions as has also been previously shown \cite{Sethia:13,Sethia:14}.
The figure shows the distribution of the $201$ oscillators in the complex plane of ($Re(W_j)\;vs\;Im(W_j)$). We observe the typical signature of an AMC in the form of a string like object representing the incoherent oscillators and a cluster (represented by a black filled square) marking the coherent oscillators which move together in the complex plane. The new feature compared to the standard AMC state \cite{Sethia:14} is the presence of another cluster (marked by a red triangle) that represents the dynamics of the dead oscillators which are no longer inactive now but have acquired a frequency and form another coherent region in the AMC. These oscillators have a lower amplitude than the originally active oscillators but oscillate at the same common collective frequency as them. Thus the inactive components of the system experience a revival due to the global coupling and modulate the coherent portion of the AMC profile.  

\begin{figure}[h!]
\begin{center}
\includegraphics[scale=0.65]{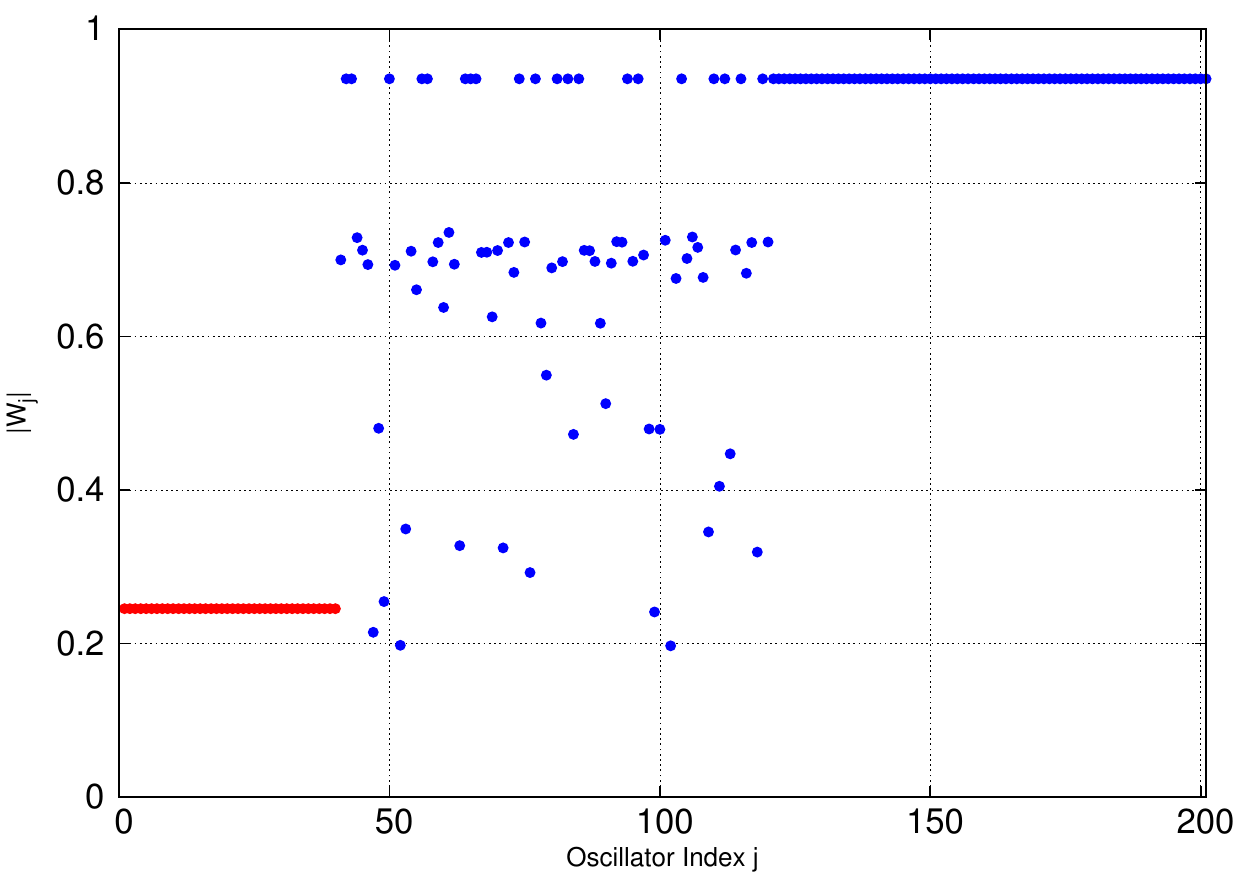}
\caption{(Color online) The chimera state is characterized by two groups of coherent oscillators that have $\mid W \mid$ around $0.2$ and $\mid W \mid$ close to $1$ respectively. The scattered points represent the incoherent part of the chimera state. The other parameters for this state are $N=201$, $K=0.7$, $C_1=-1$, $C_2 =2$, $a=b=1$ and $p=0.2$}
\label{modW}
\end{center}
\end{figure}

\begin{figure}[h!]
\begin{center}
\includegraphics[scale=0.65]{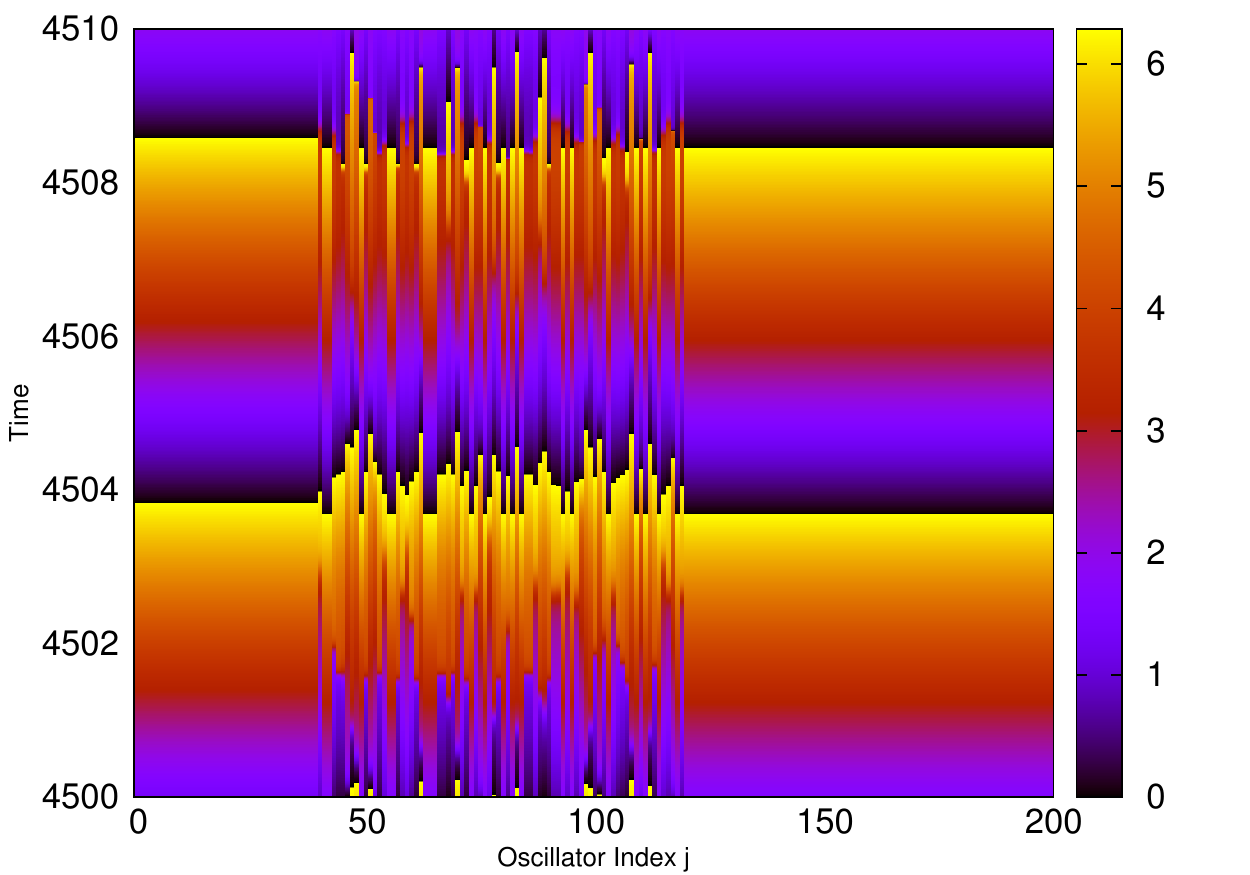}
\caption{(Color online) Phases of the oscillators for a chimera state arising from an initial state of a mixed population of active and inactive oscillators with $N = 201$, $K = 0.7$, $C_1 = -1$, $C_2 = 2$, $a = b = 1$, $p = 0.2$. }
\label{phase}
\end{center}
\end{figure}

This can be seen more clearly in the snapshots of the profiles of $\mid W_j\mid$ and phase $\phi_j$ shown in Figs.~(\ref{modW}) and (\ref{phase}) respectively for $N=201, K = 0.7, C_1 = -1, C_2 = 2$, $a = b = 1$ and $p = 0.2$. The solid line (red) close to $|W|=0.2$ in Fig.~(\ref{modW}) marks the coherent region arising from the initially inactive oscillators. They have a smaller amplitude of oscillation than the initially active ones that form the coherent region shown by the other (blue) solid line. The scattered dots show the incoherent region whose oscillators drift at different frequencies. The oscillator phases of the three regions are shown in Fig.~(\ref{phase}) where one observes that the phases of the oscillators in each coherent region remain the same and there is a finite phase difference between the two regions. The incoherent regions have a random distribution of phases among the oscillators.

\begin{figure}[h!]
\begin{center}
\includegraphics[scale=0.65]{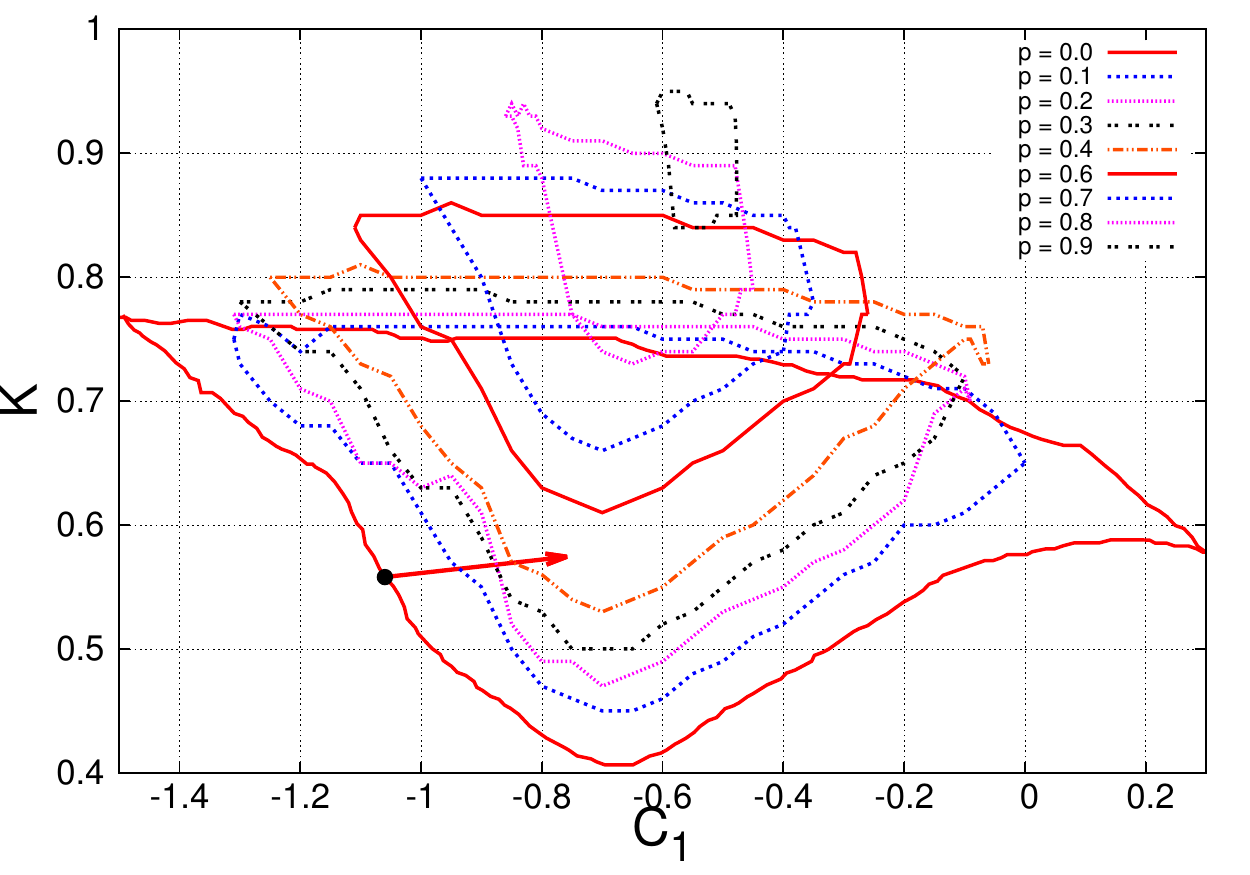}
\caption{(Color online) Phase diagram in the $C_1 - K$ space with $N = 201$, $C_2 = 2$, $a = 1$, $b = 1$ with different values of $p$. The arrow marks the shift in the values of $K$ and $C_1$ necessary to maintain an AMC state that originally existed in the $p=0$ domain to continue to remain an AMC for $p=0.1$.}
\label{C_1K}
\end{center}
\end{figure}

We have next carried out extensive numerical explorations to determine the extent and location of the parametric domain where such AMCs can occur in order to determine the changes if any from the existence region for $p=0$. Our results are shown in Fig.~(\ref{C_1K}) for several different values of $p$ where the different curves mark the outer boundaries of the existence region of AMCs for particular values of $p$. 
We find that, for a fixed value of $b=1$, as the value of $p$ is increased the parametric domain of the existence region of AMC shrinks and also shifts upward and away from the $p=0$ region towards a higher value of $K$. 
The area of this region asymptotically goes to zero as $p \rightarrow 1$. Note that there is a finite parametric domain of existence even for a $p$ value that is as large as $0.9$ indicating that AMCs can occur even in the presence of a very large number of inactive oscillators and thus are very robust to aging related changes in the system environment. 
We have also found that the revival is dependent on the value of the coupling constant $K$ as well as the parameters $p$ and $b$.

\section{Mean Field Theory}
\begin{figure}[h!]
\begin{center}
\includegraphics[scale=0.65]{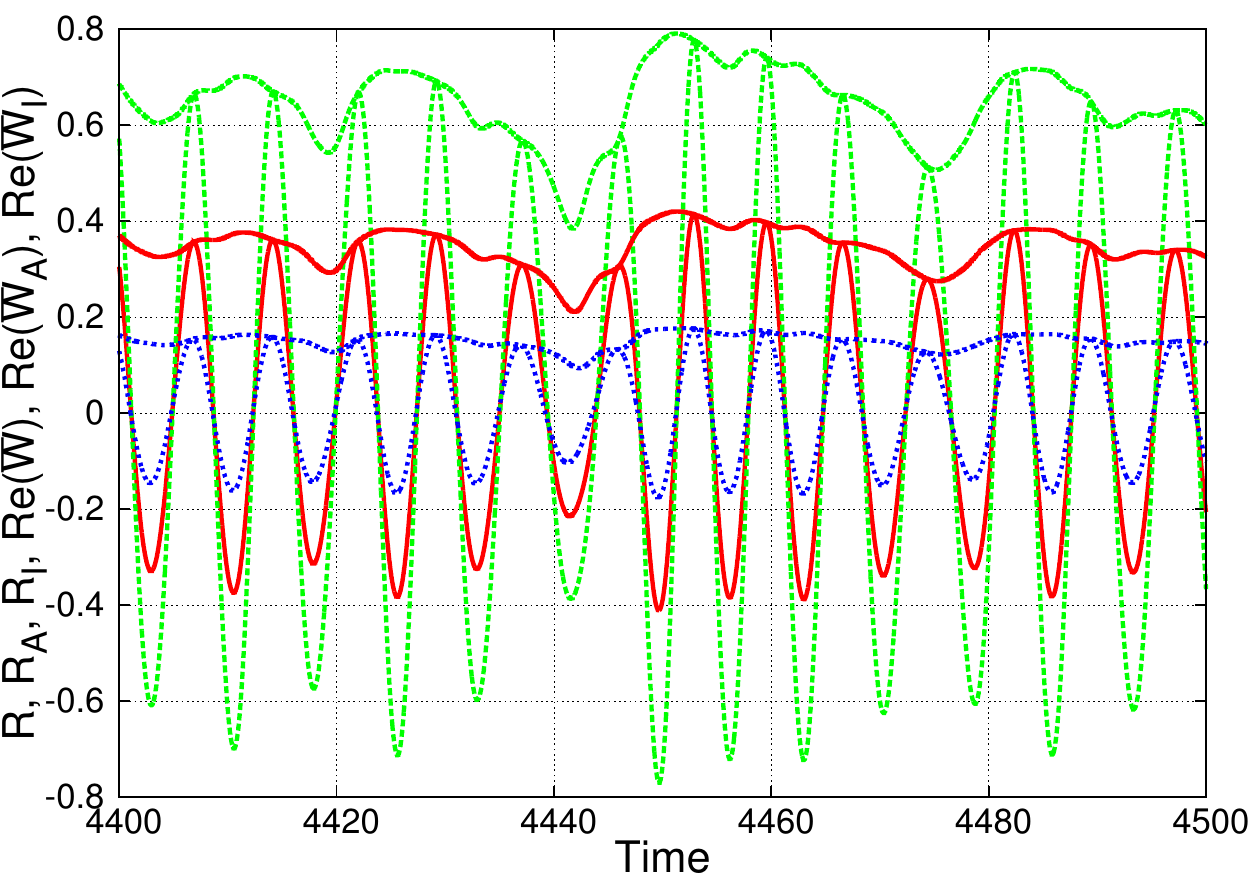}
\caption{(Color online) $R_{A}$ (green dashed line), $R_{I}$ (blue dotted line) and $R$ (red solid line) for $N = 201$, $K = 0.7$, $C_1 = -1$, $C_2 = 2$, $a = b = 1$, $p = 0.6$. The oscillatory part represents the $Re (\bar{W})$ and the line through the peaks represents $|\bar{W}|$.}
\label{meanR}
\end{center}
\end{figure}

\begin{figure}[h!]
\begin{center}
\includegraphics[scale=0.65]{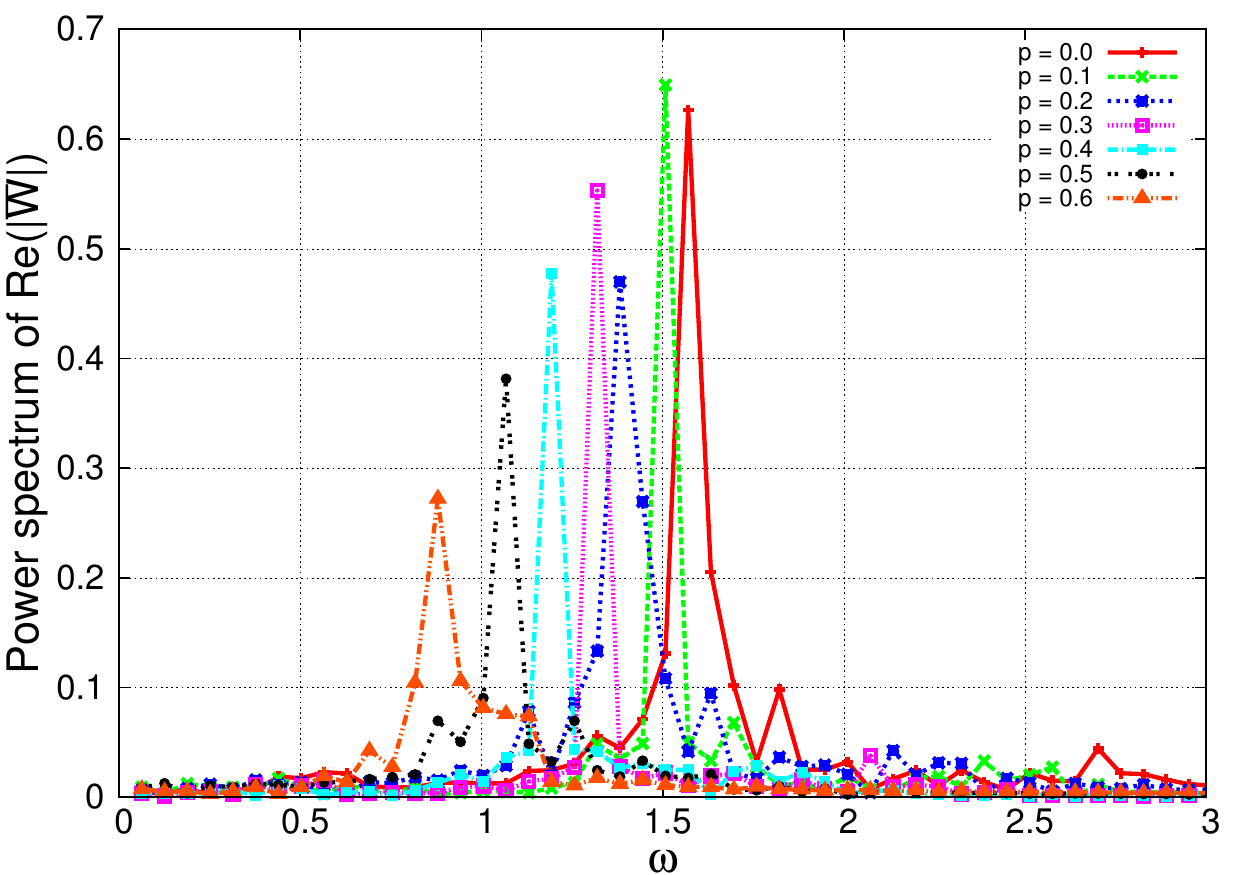}
\caption{(Color online) Power spectrum of $Re (\bar{W})$ for $N = 201$, $K = 0.7$, $C_1 = -1$, $C_2 = 2$, $a = b = 1$ with different values of $p$.}
\label{power}
\end{center}
\end{figure}

\begin{figure}[h!]
\begin{center}
\includegraphics[scale=0.65]{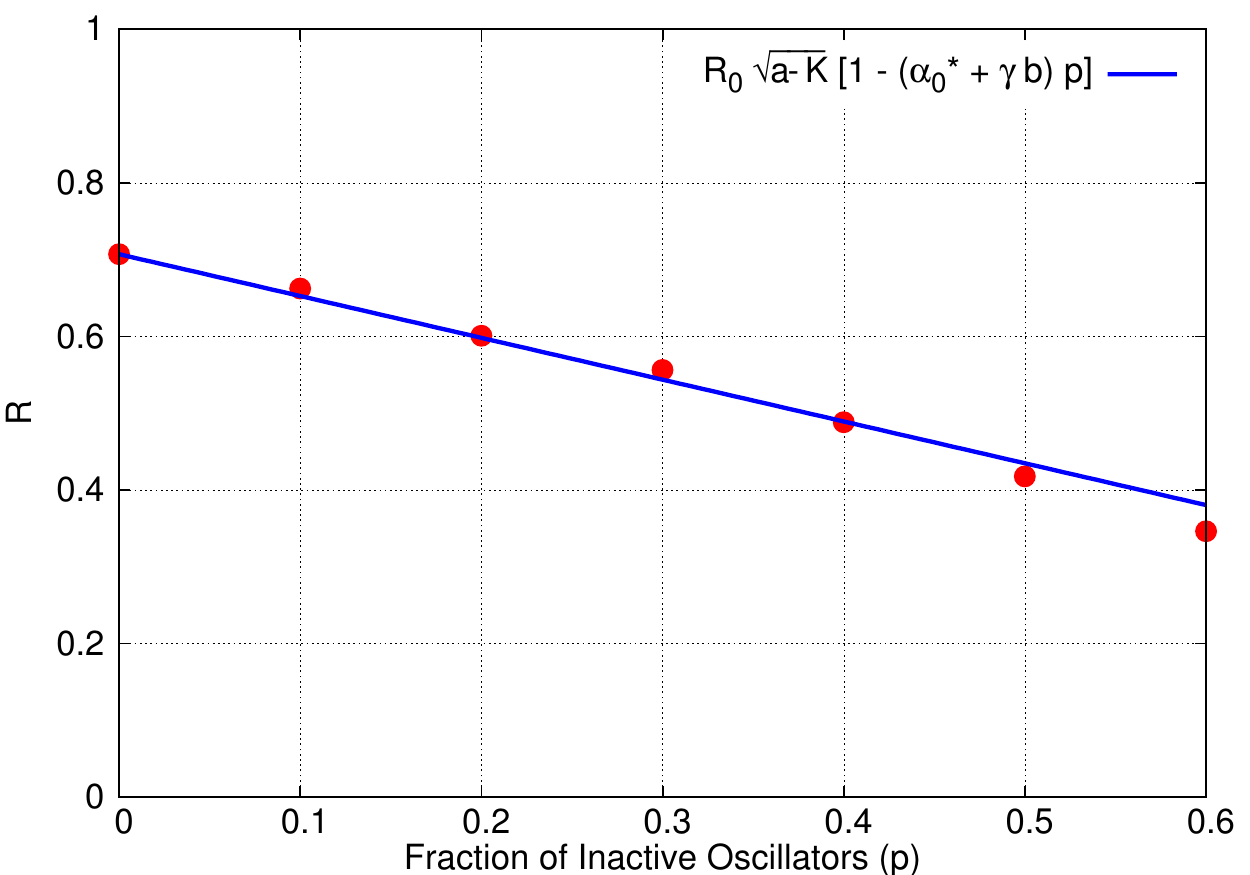}
\caption{(Color online) Variation of $R$ (red filled circle) with $p$. The blue solid line represents the fit for $R$. The parameters chosen are $N = 201$, $K = 0.7$, $C_1 = -1$, $C_2 = 2$.}
\label{fit_R}
\end{center}
\end{figure}

\begin{figure}[h!]
\begin{center}
\includegraphics[scale=0.65]{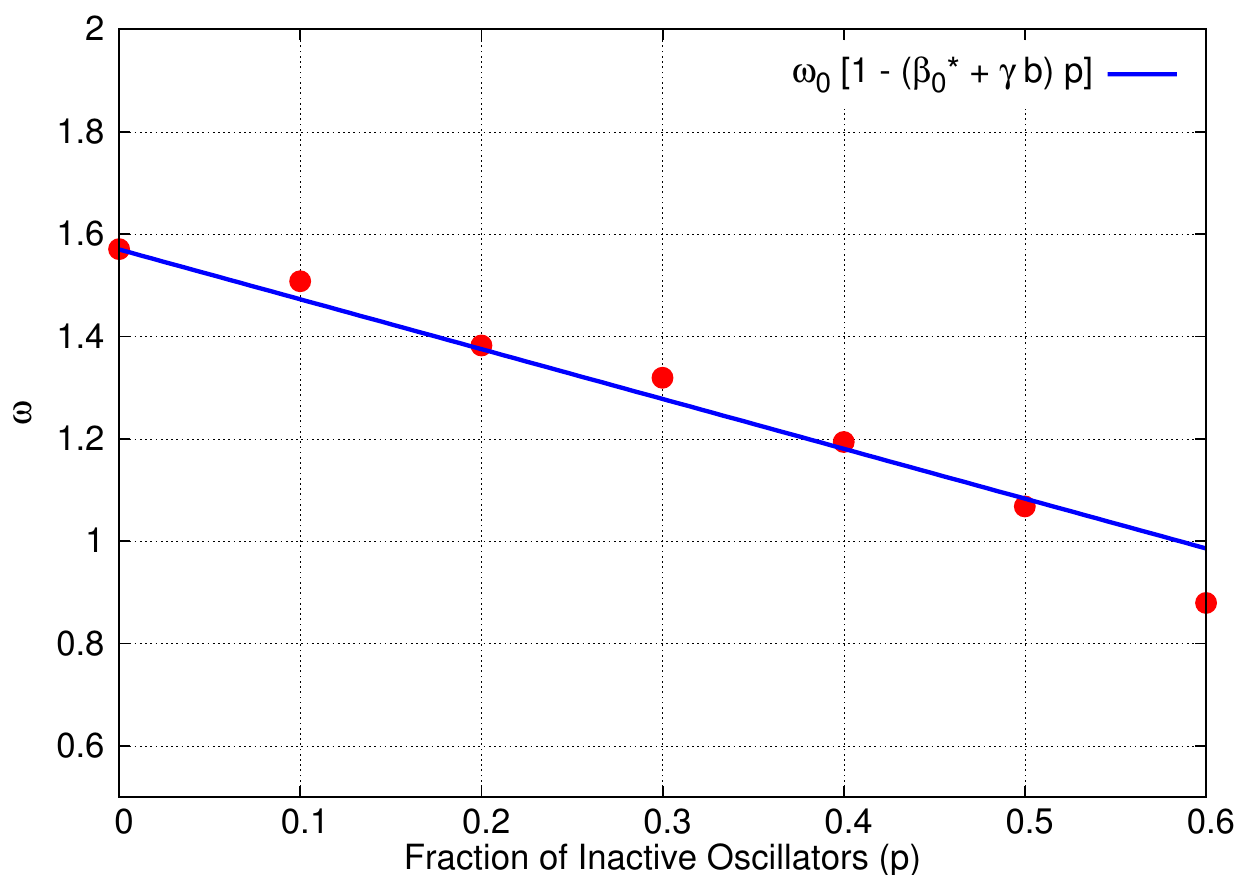}
\caption{(Color online)  Variation of $\omega$ (red filled circle) with $p$. The blue solid line represents the fit for $\omega$. The parameters chosen are $N = 201$, $K = 0.7$, $C_1 = -1$, $C_2 = 2$.}
\label{fit_w}
\end{center}
\end{figure}

In order to gain a deeper understanding of the dynamical origin of our numerical results we now analyse a  reduced model  that is a generalization of a similar system that has been employed in the past to study the chaotic and AMC states of Eq.(\ref{CCGLE}) in the absence of initially inactive oscillators \cite{Chabanol:97,Sethia:14}. We define appropriate mean field parameters 
$\overline{W}_{I}$ for the initially inactive oscillator population and $\overline{W}_{A}$ for the active population, 
as,
\begin{eqnarray}
&& \overline{W}_I(t) = \left(\frac{1}{Np} \sum\limits_{j=1}^{j=Np} W_j\right) = R_I e^{i\omega_I t}\\
&& \overline{W}_A(t) = \left(\frac{1}{N(1-p)} \sum\limits_{j=Np+1}^{j=N} W_j\right) = R_A e^{i\omega_A t}
\end{eqnarray} 
\noindent
where $R_I$ and $R_A$ are the amplitudes of these mean fields and $\omega_I$ and $\omega_A$ are their mean frequencies. 
Note that the mean field for the entire system (\ref{CCGLE}) can also be expressed as,
\begin{equation}
\overline{W}(t) = \left(\frac{1}{N} \sum\limits_{i=1}^N W_i\right) = R e^{i\omega_{T} t}\\
\end{equation}
with 
$$ R=|\overline{W}|=\frac{1}{N}\sqrt{\left(\sum\limits_{j=1}^{N} Re\left(W_{j}\right)\right)^2 + \left(\sum\limits_{j=1}^{N} Im\left(W_{j}\right)\right)^2}$$ 
The nature of $R$, $R_A$ and $R_I$ can be seen from our numerical simulation results shown in Fig.~(\ref{meanR}) where we have plotted the time evolution of $|\overline{W}|$, $|\overline{W}_I|$, $|\overline{W}_A|$, $Re(\overline{W})$, $Re(\overline{W}_A)$ and $Re(\overline{W}_I)$. We observe that the time variations of $Re(\overline{W}_I)$, $Re(\overline{W}_A)$ and $Re(\overline{W})$ are nearly periodic and identical for the active, initially inactive and the full set of oscillators. A power spectrum plot of $Re (\overline{W})$ given in Fig.~(\ref{power}) further shows that the periodicity is primarily around a single frequency, $\omega$, as denoted by the sharp peak in the power spectrum. Note also that this central frequency varies as a function of $p$ and decreases as $p$ increases. The amplitudes $R_A =|\overline{W}_{A}|, R_{I}=|\overline{W}_{I}|$ and $R =|\overline{W}|$ evolve on a slower time scale and are nearly constant with small fluctuations around a mean value. The mean values also decrease as a function of $p$ and this is shown in Fig.~(\ref{fit_R}) where the mean value of $R$ is plotted against $p$. The decrease of $\omega$ with $p$ is displayed in Fig.~(\ref{fit_w}).\\

The near constancy of the amplitudes of the mean fields and the existence of a common dominant single frequency of oscillation permits the development of a simple model system, along the lines of Ref. \cite{Sethia:14}, in terms of the driven dynamics of representative single oscillators.\\

We therefore define two single oscillator variables $B_I(t)$ and $B_A(t)$ to represent the dynamics of any one of the 
initially inactive oscillators and any one of the initially active oscillators, respectively. We take,
\begin{equation}
B_I(t) = (1-K)^{-1/2} W_j(t) e^{-i(\omega_{T} t+\phi)} \;\; (1 \leq j \leq Np)
\label{BI}
\end{equation} 
\begin{equation}
B_A(t) = (1-K)^{-1/2} W_j(t) e^{-i(\omega_{T} t+\phi)} \;\; (Np + 1 \leq j \leq N)
\label{BA}
\end{equation} 
Using (\ref{BI}) and (\ref{BA}) in (\ref{CCGLE})
and re-scaling time as
\begin{equation}
\tau = (1-K) t
\end{equation}
we get two driven single oscillator equations, 
\begin{eqnarray}
\label{reduced_eqn_I}&& \frac{\partial B_I}{\partial \tau} = \left(-\frac{1+K}{1-K} + i \Omega\right) B_I \nonumber \\
&& ~~~~~~~~~~~~ - (1+i C_2)|B_I|^2 B_I + F 
\end{eqnarray}
\begin{eqnarray}
\label{reduced_eqn_A}\frac{\partial B_A}{\partial \tau} = (1+i \Omega) B_A - (1+i C_2)|B_A|^2 B_A + F 
\end{eqnarray}
with
\begin{eqnarray}
\label{Omega} && \Omega (p) = - \frac{\omega (p) + K C_1}{1-K}\\
\label{F} && F (p) = \frac{K \sqrt{1+C_1^2}}{(1-K)^{3/2}} R (p)\\
&& C_1 = \tan(\phi), ~~~ - \frac{\pi}{2} \textless \phi \textless \frac{\pi}{2}
\end{eqnarray}
The term $F$ represents the mean field contribution of the entire set of oscillators and is a common driver for 
members of each sub-population of the oscillators. The basic difference in the dynamics of the two populations arises
from the sign of the $\alpha_j$ term, namely, $\alpha_j=-b$ for the initially inactive population and $\alpha_j=a$ for the active oscillators. We have taken $a=b=1$ to agree with our numerical simulations.
In the relations \ref{Omega} and \ref{F}, $\Omega$ and $F$ are now functions of $p$ since $R$ and $\omega$ vary with $p$ as seen from Figs.~(\ref{meanR}) and (\ref{power}) as well as from Figs.~(\ref{fit_R}) and (\ref{fit_w}).\\
The plots of Figs.~(\ref{fit_R}) and (\ref{fit_w}) also allow us to derive approximate analytic expressions to describe the dependence of $R$ and $\omega$ on $p$, namely,
\begin{eqnarray}
\label{R_scaling} && R = R_0 \sqrt{a-K} \left[1-(\alpha^\star_0+\gamma b)p\right]\\
\label{w_scaling} && \omega = \omega_0 \left[1-(\beta^\star_0+\gamma b)p\right]
\end{eqnarray} 
where for $K$ = 0.7, $C_1$ = -1, $C_2$ = 2 and $a$ = 1 we get, $R_0$ = 1.291, $|\;\omega_0\;|$ = 1.57,
$\alpha^\star_0$ = 0.63, $\beta_0^\star$ = 0.48 and $\gamma$ = 0.141. 

\begin{figure}[h!]
\begin{center}
\includegraphics[scale=0.65]{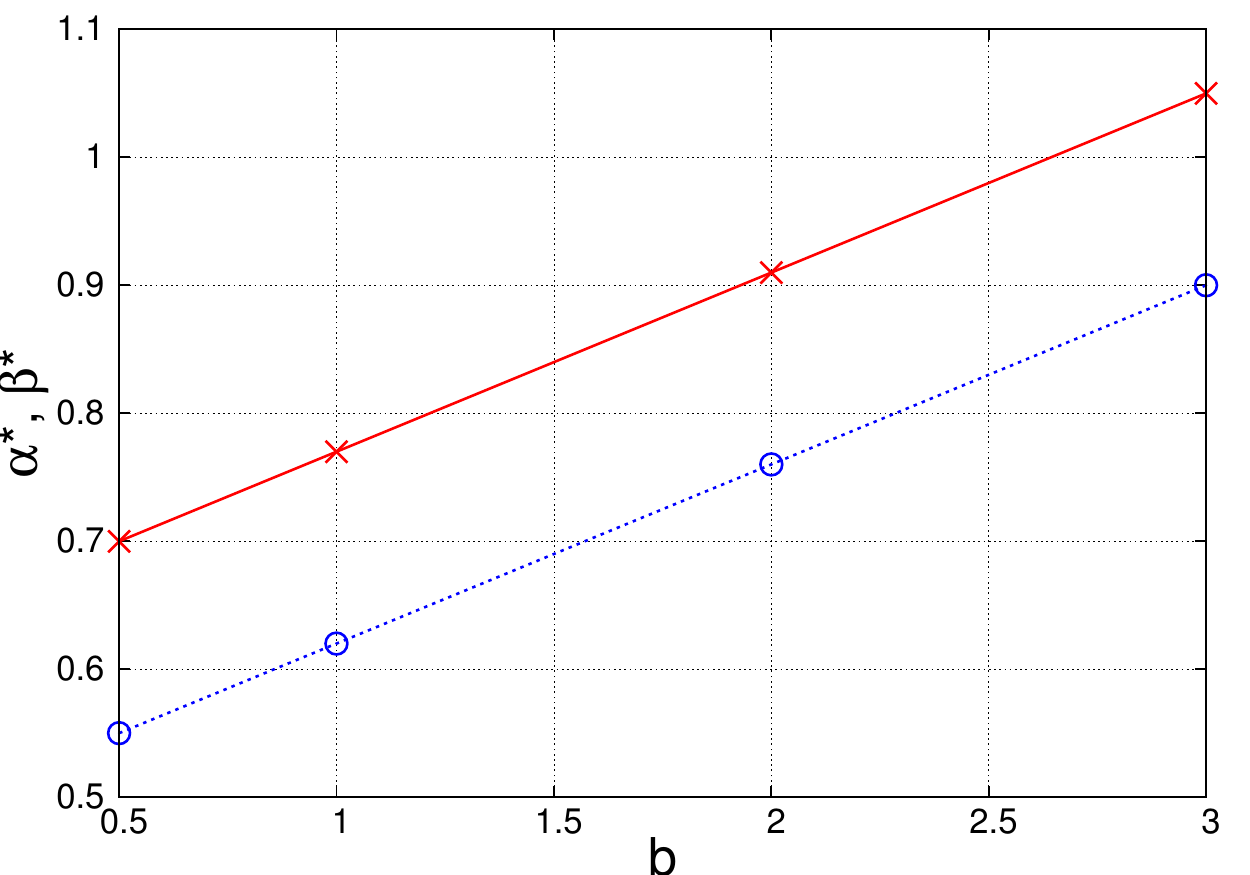}
\caption{(Color online)  Variation of $\alpha^\star$ (red solid line) and $\beta^\star$ (blue dotted line) with $b$. The parameters chosen are $N = 201$, $K = 0.7$, $C_1 = -1$, $C_2 = 2$. The linear variation indicates the relation 
$\alpha^\star = \alpha^\star_0 + \gamma b$ and $\beta^\star = \beta^\star_0 + \gamma b$}
\label{fit_b}
\end{center}
\end{figure}

\begin{figure}[h!]
\begin{center}
\includegraphics[scale=0.65]{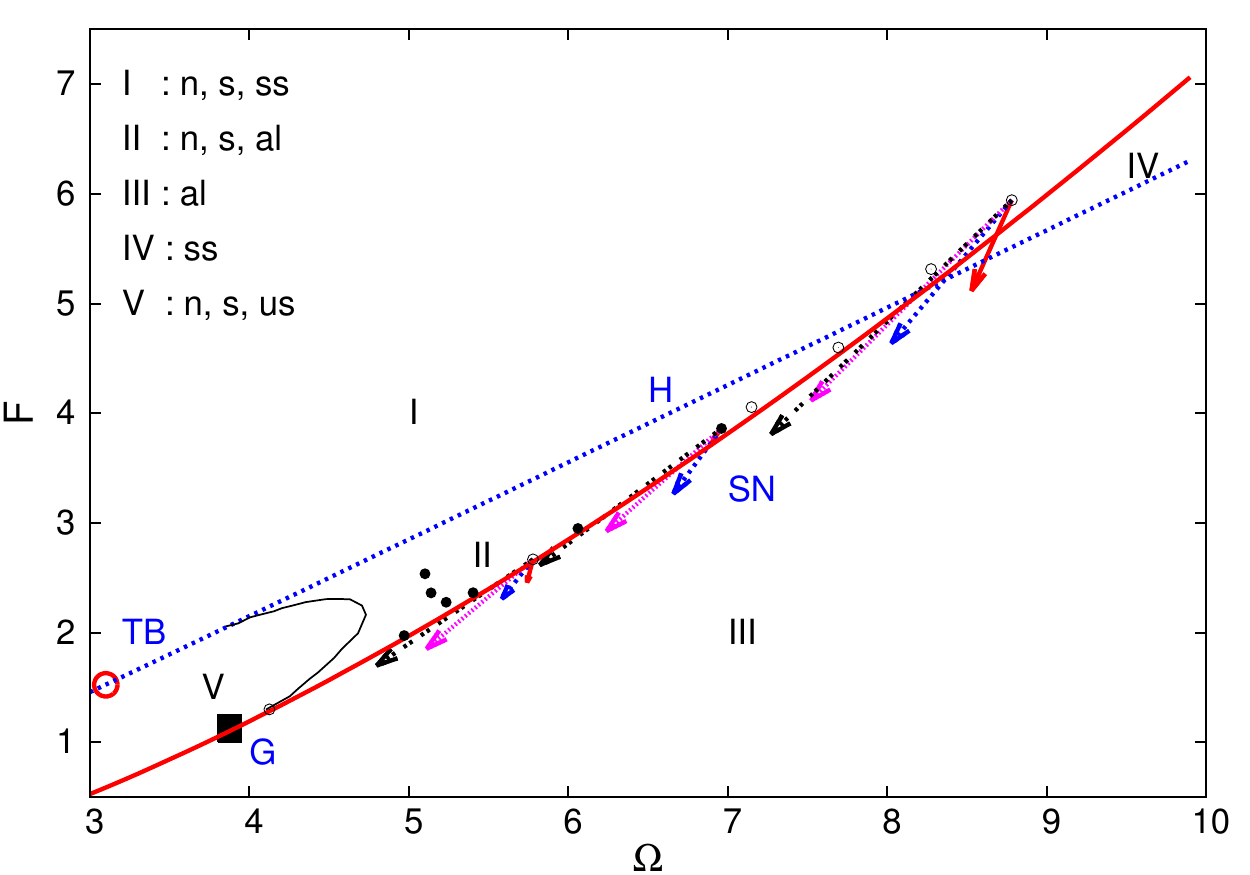}
\caption{(Color online) Bifurcation Diagram for Eq.(\ref{reduced_eqn_A}) with parameters $N = 201$, $K = 0.7$, $C_1 = -1$, $C_2 = 2$, $a = b =1$ and $p = 0$. The red solid line is the Hopf Bifurcation line (H) and the blue dotted line is the Saddle-Node Bifurcation line (SN). The thin black solid line between region II and V is the homoclinic bifurcation line. The big red circle and the black square indicate the Takens Bogdanov bifurcation point and the codimension-two point and are labeled as TB and G respectively. In the legend at the top left corner of the figure, the letter `n', `s', `al', `ss' and `us' denote a node, a saddle, an attractive limit cycle, stable spiral and unstable spiral respectively. The region between the red solid and the blue dotted lines represents the probable regime of AMC states with $p = 0$. The arrows indicate the shift of the AMC states with p = 0.1 (red solid arrow), 0.2 (blue dotted arrow), 0.3 (magenta dashed arrow), 0.4 (black dashed dotted arrow). The filled circles represent typical AMC states coexistent with synchronous states while empty circles are for AMCs in the unstable region of synchronous states.}
\label{bif_p_0}
\end{center}
\end{figure}

\begin{figure}[h!]
\begin{center}
\includegraphics[scale=0.65]{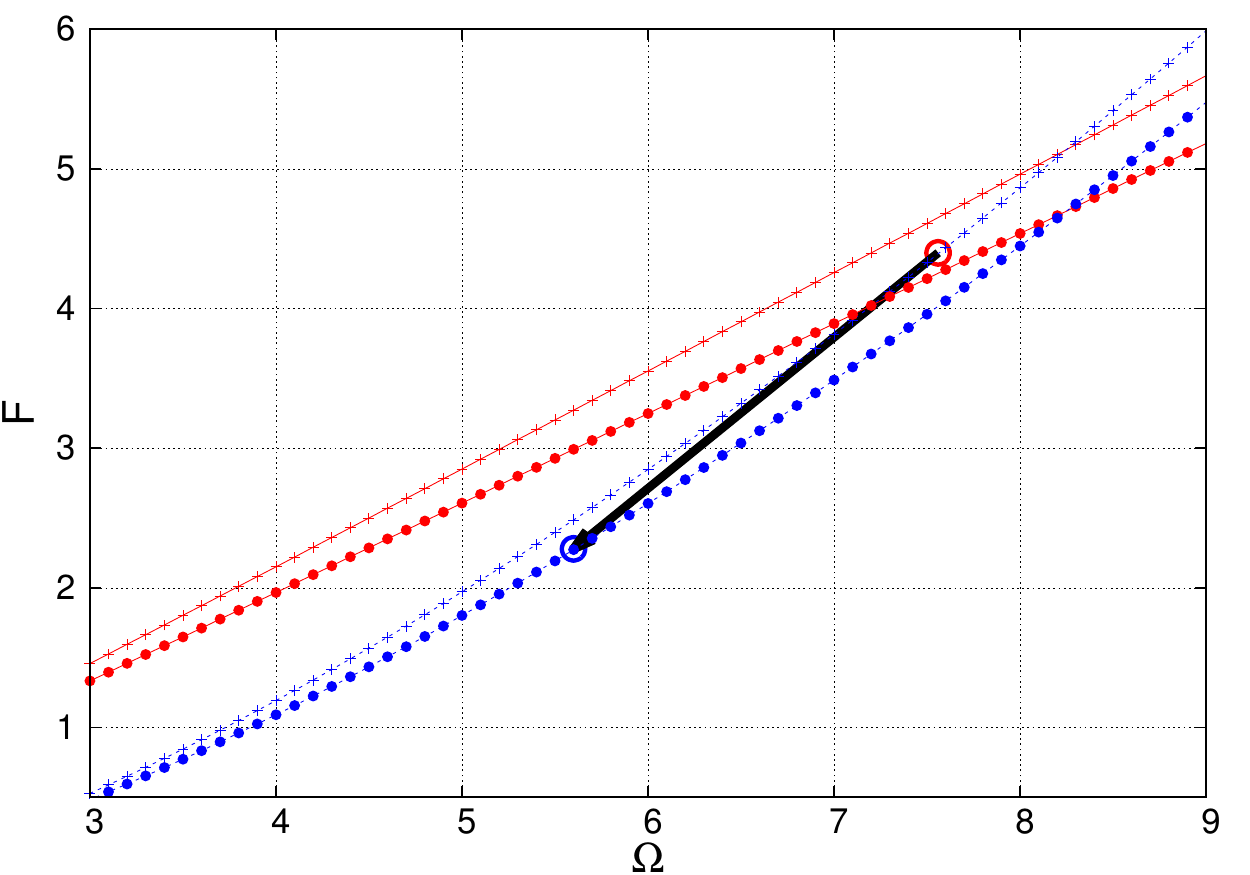}
\caption{(Color online) Bifurcation Diagram for Eq.(\ref{reduced_eqn_A}) with parameters $N = 201$, $K = 0.7$, $C_1 = -1$, $C_2 = 2$, $a = b = 1$ and $p = 0$ (``+" symbol) \& $0.6$ (filled circles). The red solid line is the Hopf Bifurcation line and the blue dotted line is the Saddle-Node Bifurcation line.}
\label{bif_p_0p6}
\end{center}
\end{figure}

We can now try to understand the dynamical origin of the behaviour of the existence region of the AMC as a function of $p$ by analyzing the two reduced equations Eq.(\ref{reduced_eqn_I} and Eq.(\ref{reduced_eqn_A}). 
For $p=0$, Eq.(\ref{reduced_eqn_A}) has been shown \cite{Chabanol:97,Sethia:14} to have a rich bifurcation diagram in the $F-\Omega$ space as shown in Fig.~(\ref{bif_p_0}) where the red (solid) line is the Hopf bifurcation line and the blue (dotted) line represents the saddle-node bifurcation line. The thin black solid line between region II and V is the homoclinic bifurcation line. The big red circle and the black square indicate the Takens Bogdanov bifurcation point and the codimension-two point and are labeled as TB and G respectively. The various regions, marked as I-V, are characterized by the existence of a single or a combination of nodes, saddle points, attractive limit cycles, stable spirals and unstable spirals. The region between the red solid and the blue dotted lines represents the probable regime of AMC states. As discussed previously \cite{Sethia:14} the AMC states arise from the coexistence of a stable node and a limit cycle or a spiral attractor close to the saddle-node curve. The fluctuations in the amplitude of the mean field then drive the oscillators towards these equilibrium points with those that go to the node constituting the coherent part of the AMC while those that populate the limit cycle or the stable spiral forming the incoherent part of the AMC. The distribution of the oscillators among these two sub-populations depends on the initial conditions and the kicks in phase space that they receive from the amplitude fluctuations. The location of these $p=0$ AMCs are marked by the open and filled circles in Fig.~ (\ref{bif_p_0}). With the increase of $p$, $\Omega$ decreases, causing the determinant of the Jacobian ($\det J$) of Eq. (\ref{reduced_eqn_A}) to monotonically decrease while keeping the trace of the Jacobian ($Tr~J$) to remain unchanged. This results in the loss of a spiral into a node leading the AMC to collapse into two or three coherent cluster states. Also, as $p$ increases, $R$ decreases and $\left[(Tr~J)^2 - 4 \det J\right]$ decreases. This leads to the loss of a node and the generation of a spiral causing the AMC states near the saddle-node bifurcation line, for $p = 0$, to evolve into chaotic states. The above two reasons account for the shrinkage of the existence region of the AMC with the increase of $p$. Furthermore due to the re-scaling of $\Omega$ and $F$ for a given value of $p$, the bifurcation plot also shifts in the $F-\Omega$ space as shown in Fig.~ (\ref{bif_p_0p6}) for $p=0.6$. For $K = 0.7$, $C_1 = -1$, $C_2 = 2$, $a = 1$, $R_0 = 1.291$, $| \omega_0 | = 1.57$, $\alpha^\star_0 = 0.63$, $\beta^\star_0 = 0.48$, $\gamma = 0.141$ and $p = 0$, we find $\Omega = 7.56$ and $F = 4.259$ while for $p=0.6$ we get $\Omega = 5.6$ and $F = 2.28$. This shift leads to a relocation of the position of the new AMC state as indicated by the black arrow in the figure. The amount of shift can also be estimated by using relations (\ref{Omega} - \ref{w_scaling}). In Fig (\ref{bif_p_0}) we show such shifts for various values of $p$ by the directions and lengths of arrows originating from various points of the bifurcation diagram for $p=0$.

We can also use expressions (\ref{Omega}) and (\ref{F}) to understand the shift in the existence domain of the AMC in the $C_1-K$ space, as shown in Fig.~(\ref{C_1K}), by a simple approximate analysis for small values of $p$. We seek the shifts in the values of $K$ and $C_1$ for which a given AMC for $p=0$ will still remain an AMC for a finite value of $p$. This can be obtained by doing a linear perturbation analysis around the values of $K$,$C_1$, $\omega$ and $R$ for $p=0$ in expressions (\ref{Omega}) and (\ref{F}). Writing,
$$ \omega = \omega_0 + \delta \omega \;\;;\;\; R =R_0 + \delta R $$
$$ K = K_0 + \delta K \;\;;\;\; C_1 = C_{1_0} + \delta C_1 $$ 
(where the subscript $0$ labels values at $p=0$ and the terms with $\delta$ represent small perturbations) substituting in (\ref{Omega}) and (\ref{F}) and retaining only the linear terms of the perturbed quantities, we get,
\begin{eqnarray}
&& \delta C_1 = \frac{1}{K_0} \left(\Omega_0 ~ \delta K - \delta \omega - \delta K ~ C_{1_0} \right)\\
&& \left[3F_0^2 (1-K_0)^2-2K_0 R_0^2(1+C_{1_0}\Omega_0) \right] \delta K \nonumber\\
&& ~~~ = - 2K_0 R_0 [K_0~\delta R + K_0 C_{1_0}^2~\delta R - C_{1_0} R_0 ~ \delta \omega]
\end{eqnarray}
The above equations can be solved for $\delta K$ and $\delta C_1$ using the numerically obtained values for $\delta \omega$, $\delta R$ for a chosen small value of $p$. For $p=0.1$ we have $\delta \omega =-0.12$ and $\delta R = -0.06$. Using the above values at $C_{1_0} =-0.7$ and $K_0=0.7$ we get $\delta K = 0.004$ and $\delta C_1 = 0.18$.  These shifts in the values of $K$ and $C_1$ are depicted by an arrow in Fig.~(\ref{C_1K}) and indicate the shift in the location of an AMC in the $K\;-\;C_1$ space.
The direction of the shift agrees quite well with the observed shift in the domain of the $p=0.1$ AMC compared to the domain with $p=0$. \\

\begin{figure}[h!]
\begin{center}
a)\includegraphics[scale=0.65]{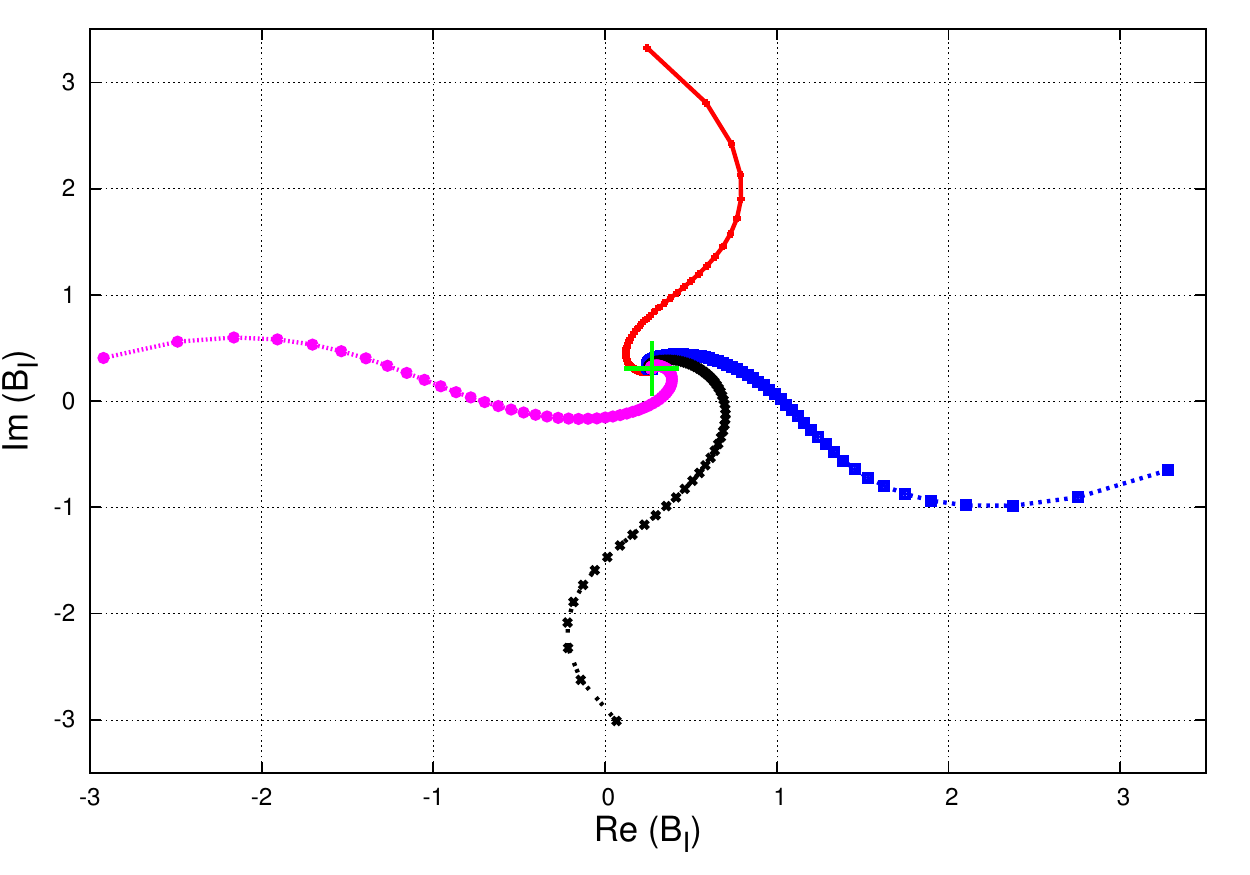}
b)\includegraphics[scale=0.65]{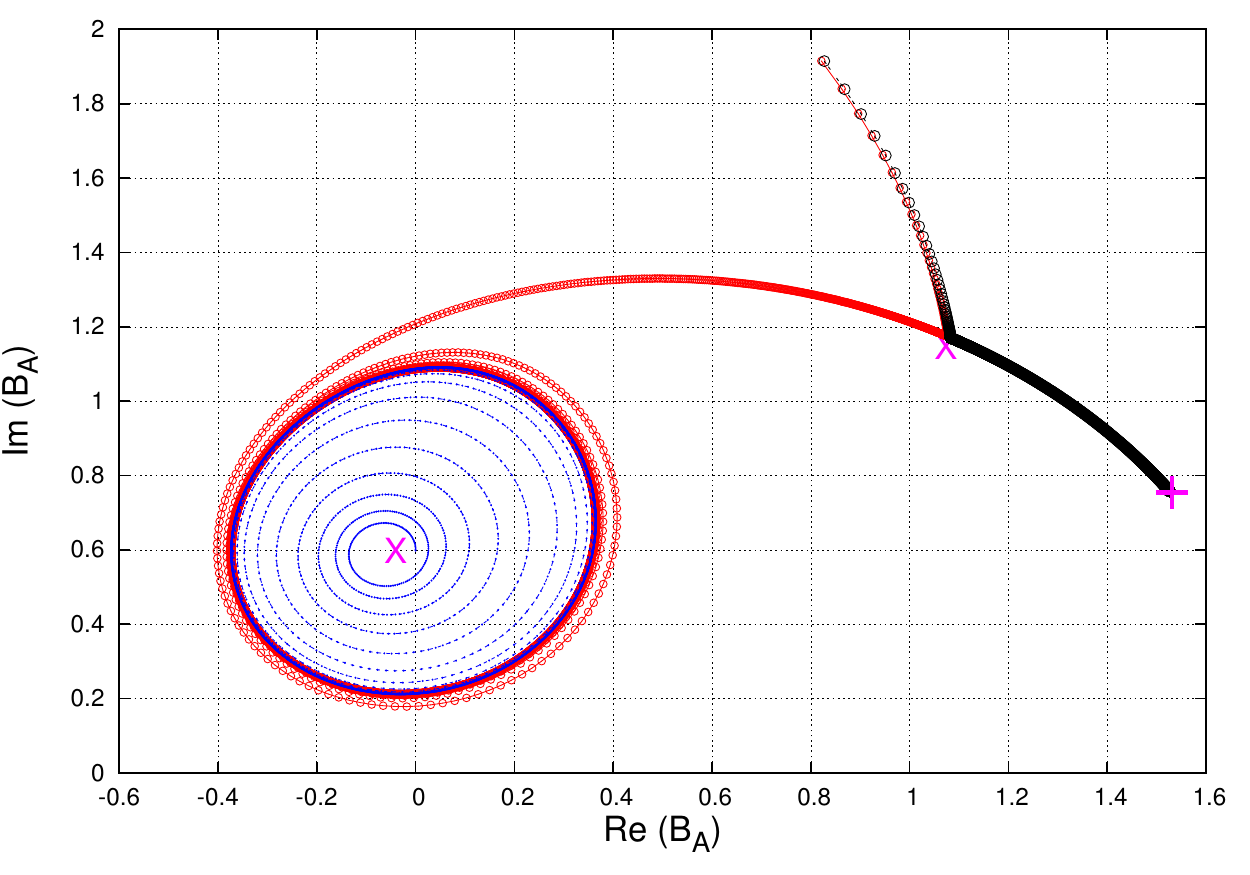}
\caption{(Color online) (a) Phase space of $Re(B_I)\;-\;Im(B_I)$ for Eq.(\ref{reduced_eqn_I}) with parameters $N = 201$, $K = 0.7$, $C_1 = -1$, $C_2 = 2$, $a = b = 1$ and $p = 0.2$. (b) Phase space of $Re(B_A)\;-\;Im(B_A)$ for Eq.(\ref{reduced_eqn_A}) with parameters $N = 201$, $K = 0.7$, $C_1 = -1$, $C_2 = 2$, $a = b = 1$ and $p = 0.2$. }
\label{reduced_BA_BI}
\end{center}
\end{figure}

We next turn to an analysis of Eq.(\ref{reduced_eqn_I}) to understand the behavior of the sub-population of initially inactive oscillators. In contrast to Eq.(\ref{reduced_eqn_A}), this equation has a very simple topological structure in that it only admits a stable fixed point  which corresponds to a periodic motion (with frequency $\omega$) of the corresponding variable $W_j$. This explains the existence of the coherent region marked in red (around $|W|\approx 0.2$) shown in Fig.~(\ref{modW}). In Fig.~\ref{reduced_BA_BI}(a) we plot the evolution dynamics of Eq.(\ref{reduced_eqn_I}) in the phase space of $(Re(B_I)-Im(B_I))$ for $K = 0.7$, $C_1 = -1$, $C_2 = 2$ and $p = 0.2$ with $R = 0.6009$ and $|\omega| = 1.3823$. The location of the fixed point at 
$Re(B_I) = 0.272, Im(B_I) = 0.3078$ corresponds to $|W_I|= |B_I|\sqrt{1-K} =  0.2681$, which corresponds to the left coherent cluster shown in Fig.~(\ref{modW}). For a comparison we also show in Fig.~\ref{reduced_BA_BI}(b) a corresponding phase diagram for Eq.~(\ref{reduced_eqn_A}) for the same set of parameters as Fig.~\ref{reduced_BA_BI}(a). We see the existence of three fixed points - one stable (marked with a $+$ symbol) and two unstable (marked with a $\times$ symbol) and a limit cycle - the ingredients for the creation of an AMC. Thus the combined  dynamics of the two model equations (\ref{reduced_eqn_I}) and (\ref{reduced_eqn_A}) provide a composite picture of the existence domain and characteristic features of the AMC in the presence of a population of inactive oscillators.

\section{Summary and Discussion}
To summarize, we have studied the influence of a population of inactive oscillators on the formation and dynamical features of amplitude mediated chimera states in an ensemble of globally coupled Ginzburg-Landau oscillators. From our numerical investigations we find that the inactive oscillators influence the AMCs in several distinct ways. The coupling with the rest of the active oscillators revives their oscillatory properties and they become a part of the AMC as a separate coherent cluster thereby modulating the structure of the AMC. Their presence also reduces the overall frequency of the coherent group of oscillators. Finally they shrink the existence region of the AMCs in the parametric space of the coupling strength $K$ and the constant $C_1$ (where $KC_1$ is the imaginary component of the coupling constant). This region continously shrinks and shifts in this parametric domain as a function of $p$. Remarkably, the AMCs continue to exist (albeit in a very small parametric domain) even for $p$ as large as $0.9$ which is indicative of their robustness to aging effects in the system. Our numerical results can be well understood from an analytic study of a reduced model that is derived from a mean field theory and consists of two driven nonlinear oscillators that are representative of typical members of the sub-populations of initially inactive oscillators and the active oscillators. The driving term for both these single oscillators equations is the mean field arising from the global coupling of all the oscillators. A bifurcation analysis of both these model equations 
provides a good qualitative understanding of the changes taking place in the existence domain of the AMCs due to the presence of the inactive oscillators. AMCs are a generalized class of chimera states in which both amplitude and phase variations of the oscillators are retained and their existence is not constrained by the need to have non-local forms of coupling. Our findings can therefore have a wider applicability and be practically relevant for such states in biological or physical systems where aging can diminish the functional abilities of component parts. 

\section{Acknowledgement}
R.M. thanks K. Premalatha at Bharathidasan University, India for simulation related help and Pritibhajan Byakti at Indian Association of Cultivation of Science, Kolkata and Udaya Maurya at Institute for Plasma Research, India for help on Mathematica. He also acknowledges valuable discussions with Bhumika Thakur, P N Maya, Mrityunjay Kundu and Gautam C Sethia at Institute for Plasma Research, India. The authors thank two anonymous referees for several insightful comments and suggestions.



\bibliography{biblio}

\begin{thebibliography}{30}
\expandafter\ifx\csname natexlab\endcsname\relax\def\natexlab#1{#1}\fi
\expandafter\ifx\csname bibnamefont\endcsname\relax
  \def\bibnamefont#1{#1}\fi
\expandafter\ifx\csname bibfnamefont\endcsname\relax
  \def\bibfnamefont#1{#1}\fi
\expandafter\ifx\csname citenamefont\endcsname\relax
  \def\citenamefont#1{#1}\fi
\expandafter\ifx\csname url\endcsname\relax
  \def\url#1{\texttt{#1}}\fi
\expandafter\ifx\csname urlprefix\endcsname\relax\def\urlprefix{URL }\fi
\providecommand{\bibinfo}[2]{#2}
\providecommand{\eprint}[2][]{\url{#2}}

\bibitem[{\citenamefont{Motter}(2010)}]{Motter:10}
\bibinfo{author}{\bibfnamefont{A.~E.} \bibnamefont{Motter}},
  \bibinfo{journal}{Nat. Phys.} \textbf{\bibinfo{volume}{6}},
  \bibinfo{pages}{164} (\bibinfo{year}{2010}).

\bibitem[{\citenamefont{Panaggio and Abrams}(2015)}]{Panaggio:15}
\bibinfo{author}{\bibfnamefont{M.~J.} \bibnamefont{Panaggio}} \bibnamefont{and}
  \bibinfo{author}{\bibfnamefont{D.~M.} \bibnamefont{Abrams}},
  \bibinfo{journal}{Nonlinearity} \textbf{\bibinfo{volume}{28}},
  \bibinfo{pages}{R67} (\bibinfo{year}{2015}),
  \urlprefix\url{http://stacks.iop.org/0951-7715/28/i=3/a=R67}.

\bibitem[{\citenamefont{Kuramoto and Battogtokh}(2002)}]{Kuramoto:02}
\bibinfo{author}{\bibfnamefont{Y.}~\bibnamefont{Kuramoto}} \bibnamefont{and}
  \bibinfo{author}{\bibfnamefont{D.}~\bibnamefont{Battogtokh}},
  \bibinfo{journal}{Nonlin. Phenom. Compl. Sys.} \textbf{\bibinfo{volume}{5}},
  \bibinfo{pages}{380} (\bibinfo{year}{2002}).

\bibitem[{\citenamefont{Sethia et~al.}(2013)\citenamefont{Sethia, Sen, and
  Johnston}}]{Sethia:13}
\bibinfo{author}{\bibfnamefont{G.~C.} \bibnamefont{Sethia}},
  \bibinfo{author}{\bibfnamefont{A.}~\bibnamefont{Sen}}, \bibnamefont{and}
  \bibinfo{author}{\bibfnamefont{G.~L.} \bibnamefont{Johnston}},
  \bibinfo{journal}{Phys. Rev. E} \textbf{\bibinfo{volume}{88}},
  \bibinfo{pages}{042917} (\bibinfo{year}{2013}),
  \urlprefix\url{https://link.aps.org/doi/10.1103/PhysRevE.88.042917}.

\bibitem[{\citenamefont{Sethia and Sen}(2014)}]{Sethia:14}
\bibinfo{author}{\bibfnamefont{G.~C.} \bibnamefont{Sethia}} \bibnamefont{and}
  \bibinfo{author}{\bibfnamefont{A.}~\bibnamefont{Sen}},
  \bibinfo{journal}{Physical Review Letters} \textbf{\bibinfo{volume}{112}},
  \bibinfo{pages}{144101} (\bibinfo{year}{2014}).

\bibitem[{\citenamefont{Santos et~al.}(2017)\citenamefont{Santos, Szezech,
  Borges, Iarosz, Caldas, Batista, Viana, and Kurths}}]{Santos:17}
\bibinfo{author}{\bibfnamefont{M.}~\bibnamefont{Santos}},
  \bibinfo{author}{\bibfnamefont{J.}~\bibnamefont{Szezech}},
  \bibinfo{author}{\bibfnamefont{F.}~\bibnamefont{Borges}},
  \bibinfo{author}{\bibfnamefont{K.}~\bibnamefont{Iarosz}},
  \bibinfo{author}{\bibfnamefont{I.}~\bibnamefont{Caldas}},
  \bibinfo{author}{\bibfnamefont{A.}~\bibnamefont{Batista}},
  \bibinfo{author}{\bibfnamefont{R.}~\bibnamefont{Viana}}, \bibnamefont{and}
  \bibinfo{author}{\bibfnamefont{J.}~\bibnamefont{Kurths}},
  \bibinfo{journal}{Chaos, Solitons \& Fractals}
  \textbf{\bibinfo{volume}{101}}, \bibinfo{pages}{86 } (\bibinfo{year}{2017}),
  ISSN \bibinfo{issn}{0960-0779},
  \urlprefix\url{http://www.sciencedirect.com/science/article/pii/S0960077917302230}.

\bibitem[{\citenamefont{Montbri\'o et~al.}(2004)\citenamefont{Montbri\'o,
  Kurths, and Blasius}}]{Montbrio:04}
\bibinfo{author}{\bibfnamefont{E.}~\bibnamefont{Montbri\'o}},
  \bibinfo{author}{\bibfnamefont{J.}~\bibnamefont{Kurths}}, \bibnamefont{and}
  \bibinfo{author}{\bibfnamefont{B.}~\bibnamefont{Blasius}},
  \bibinfo{journal}{Phys. Rev. E} \textbf{\bibinfo{volume}{70}},
  \bibinfo{pages}{056125} (\bibinfo{year}{2004}),
  \urlprefix\url{https://link.aps.org/doi/10.1103/PhysRevE.70.056125}.

\bibitem[{\citenamefont{Laing}(2009)}]{Laing:09}
\bibinfo{author}{\bibfnamefont{C.~R.} \bibnamefont{Laing}},
  \bibinfo{journal}{Chaos: An Interdisciplinary Journal of Nonlinear Science}
  \textbf{\bibinfo{volume}{19}}, \bibinfo{pages}{013113}
  (\bibinfo{year}{2009}), \eprint{http://dx.doi.org/10.1063/1.3068353},
  \urlprefix\url{http://dx.doi.org/10.1063/1.3068353}.

\bibitem[{\citenamefont{Sethia et~al.}(2008)\citenamefont{Sethia, Sen, and
  Atay}}]{Sethia:08}
\bibinfo{author}{\bibfnamefont{G.~C.} \bibnamefont{Sethia}},
  \bibinfo{author}{\bibfnamefont{A.}~\bibnamefont{Sen}}, \bibnamefont{and}
  \bibinfo{author}{\bibfnamefont{F.~M.} \bibnamefont{Atay}},
  \bibinfo{journal}{Phys. Rev. Lett.} \textbf{\bibinfo{volume}{100}},
  \bibinfo{pages}{144102} (\bibinfo{year}{2008}).

\bibitem[{\citenamefont{Tinsley et~al.}(2012)\citenamefont{Tinsley, Nkomo, and
  Showalter}}]{Tinsley:12}
\bibinfo{author}{\bibfnamefont{M.~R.} \bibnamefont{Tinsley}},
  \bibinfo{author}{\bibfnamefont{S.}~\bibnamefont{Nkomo}}, \bibnamefont{and}
  \bibinfo{author}{\bibfnamefont{K.}~\bibnamefont{Showalter}},
  \bibinfo{journal}{Nature Physics} \textbf{\bibinfo{volume}{8}},
  \bibinfo{pages}{662} (\bibinfo{year}{2012}).

\bibitem[{\citenamefont{Nkomo et~al.}(2013)\citenamefont{Nkomo, Tinsley, and
  Showalter}}]{Nkomo:13}
\bibinfo{author}{\bibfnamefont{S.}~\bibnamefont{Nkomo}},
  \bibinfo{author}{\bibfnamefont{M.~R.} \bibnamefont{Tinsley}},
  \bibnamefont{and}
  \bibinfo{author}{\bibfnamefont{K.}~\bibnamefont{Showalter}},
  \bibinfo{journal}{Physical Review letters} \textbf{\bibinfo{volume}{110}},
  \bibinfo{pages}{244102} (\bibinfo{year}{2013}).

\bibitem[{\citenamefont{Hagerstrom et~al.}(2012)\citenamefont{Hagerstrom,
  Murphy, Roy, H{\"o}vel, Omelchenko, and Sch{\"o}ll}}]{Hagerstrom:12}
\bibinfo{author}{\bibfnamefont{A.~M.} \bibnamefont{Hagerstrom}},
  \bibinfo{author}{\bibfnamefont{T.~E.} \bibnamefont{Murphy}},
  \bibinfo{author}{\bibfnamefont{R.}~\bibnamefont{Roy}},
  \bibinfo{author}{\bibfnamefont{P.}~\bibnamefont{H{\"o}vel}},
  \bibinfo{author}{\bibfnamefont{I.}~\bibnamefont{Omelchenko}},
  \bibnamefont{and}
  \bibinfo{author}{\bibfnamefont{E.}~\bibnamefont{Sch{\"o}ll}},
  \bibinfo{journal}{Nature Physics} \textbf{\bibinfo{volume}{8}},
  \bibinfo{pages}{658} (\bibinfo{year}{2012}).

\bibitem[{\citenamefont{Martens et~al.}(2013)\citenamefont{Martens, Thutupalli,
  Fourri{\`e}re, and Hallatschek}}]{Martens:13}
\bibinfo{author}{\bibfnamefont{E.~A.} \bibnamefont{Martens}},
  \bibinfo{author}{\bibfnamefont{S.}~\bibnamefont{Thutupalli}},
  \bibinfo{author}{\bibfnamefont{A.}~\bibnamefont{Fourri{\`e}re}},
  \bibnamefont{and}
  \bibinfo{author}{\bibfnamefont{O.}~\bibnamefont{Hallatschek}},
  \bibinfo{journal}{Proceedings of the National Academy of Sciences}
  \textbf{\bibinfo{volume}{110}}, \bibinfo{pages}{10563}
  (\bibinfo{year}{2013}).

\bibitem[{\citenamefont{Gambuzza et~al.}(2014)\citenamefont{Gambuzza,
  Buscarino, Chessari, Fortuna, Meucci, and Frasca}}]{Gambuzza:14}
\bibinfo{author}{\bibfnamefont{L.~V.} \bibnamefont{Gambuzza}},
  \bibinfo{author}{\bibfnamefont{A.}~\bibnamefont{Buscarino}},
  \bibinfo{author}{\bibfnamefont{S.}~\bibnamefont{Chessari}},
  \bibinfo{author}{\bibfnamefont{L.}~\bibnamefont{Fortuna}},
  \bibinfo{author}{\bibfnamefont{R.}~\bibnamefont{Meucci}}, \bibnamefont{and}
  \bibinfo{author}{\bibfnamefont{M.}~\bibnamefont{Frasca}},
  \bibinfo{journal}{Phys. Rev. E} \textbf{\bibinfo{volume}{90}},
  \bibinfo{pages}{032905} (\bibinfo{year}{2014}),
  \urlprefix\url{https://link.aps.org/doi/10.1103/PhysRevE.90.032905}.

\bibitem[{\citenamefont{Wickramasinghe and Kiss}(2013)}]{Wickramasinghe:13}
\bibinfo{author}{\bibfnamefont{M.}~\bibnamefont{Wickramasinghe}}
  \bibnamefont{and} \bibinfo{author}{\bibfnamefont{I.~Z.} \bibnamefont{Kiss}},
  \bibinfo{journal}{PloS one} \textbf{\bibinfo{volume}{8}},
  \bibinfo{pages}{e80586} (\bibinfo{year}{2013}).

\bibitem[{\citenamefont{Abrams et~al.}(2008)\citenamefont{Abrams, Mirollo,
  Strogatz, and Wiley}}]{Abrams:08}
\bibinfo{author}{\bibfnamefont{D.~M.} \bibnamefont{Abrams}},
  \bibinfo{author}{\bibfnamefont{R.}~\bibnamefont{Mirollo}},
  \bibinfo{author}{\bibfnamefont{S.~H.} \bibnamefont{Strogatz}},
  \bibnamefont{and} \bibinfo{author}{\bibfnamefont{D.~A.} \bibnamefont{Wiley}},
  \bibinfo{journal}{Phys. Rev. Lett} \textbf{\bibinfo{volume}{101}},
  \bibinfo{pages}{084103} (\bibinfo{year}{2008}).

\bibitem[{\citenamefont{Rattenborg et~al.}(2000)\citenamefont{Rattenborg,
  Amlaner, and Lima}}]{Rattenborg:00}
\bibinfo{author}{\bibfnamefont{N.~C.} \bibnamefont{Rattenborg}},
  \bibinfo{author}{\bibfnamefont{C.}~\bibnamefont{Amlaner}}, \bibnamefont{and}
  \bibinfo{author}{\bibfnamefont{S.}~\bibnamefont{Lima}},
  \bibinfo{journal}{Neuroscience \& Biobehavioral Reviews}
  \textbf{\bibinfo{volume}{24}}, \bibinfo{pages}{817} (\bibinfo{year}{2000}).

\bibitem[{\citenamefont{Shanahan}(2010)}]{Shanahan:10}
\bibinfo{author}{\bibfnamefont{M.}~\bibnamefont{Shanahan}},
  \bibinfo{journal}{Chaos: An Interdisciplinary Journal of Nonlinear Science}
  \textbf{\bibinfo{volume}{20}}, \bibinfo{pages}{013108}
  (\bibinfo{year}{2010}), \eprint{http://dx.doi.org/10.1063/1.3305451},
  \urlprefix\url{http://dx.doi.org/10.1063/1.3305451}.

\bibitem[{\citenamefont{Wildie and Shanahan}(2012)}]{Wildie:12}
\bibinfo{author}{\bibfnamefont{M.}~\bibnamefont{Wildie}} \bibnamefont{and}
  \bibinfo{author}{\bibfnamefont{M.}~\bibnamefont{Shanahan}},
  \bibinfo{journal}{Chaos: An Interdisciplinary Journal of Nonlinear Science}
  \textbf{\bibinfo{volume}{22}}, \bibinfo{pages}{043131}
  (\bibinfo{year}{2012}), \eprint{http://dx.doi.org/10.1063/1.4766592},
  \urlprefix\url{http://dx.doi.org/10.1063/1.4766592}.

\bibitem[{\citenamefont{Majhi et~al.}(2016)\citenamefont{Majhi, Perc, , and
  Ghosh}}]{Majhi:16}
\bibinfo{author}{\bibfnamefont{S.}~\bibnamefont{Majhi}},
  \bibinfo{author}{\bibfnamefont{M.}~\bibnamefont{Perc}}, , \bibnamefont{and}
  \bibinfo{author}{\bibfnamefont{D.}~\bibnamefont{Ghosh}},
  \bibinfo{journal}{Scientific Reports} \textbf{\bibinfo{volume}{6}},
  \bibinfo{pages}{39033} (\bibinfo{year}{2016}).

\bibitem[{\citenamefont{Naik et~al.}(2017)\citenamefont{Naik, Banerjee, Bapi,
  Deco, and Roy}}]{Naik:17}
\bibinfo{author}{\bibfnamefont{S.}~\bibnamefont{Naik}},
  \bibinfo{author}{\bibfnamefont{A.}~\bibnamefont{Banerjee}},
  \bibinfo{author}{\bibfnamefont{R.~S.} \bibnamefont{Bapi}},
  \bibinfo{author}{\bibfnamefont{G.}~\bibnamefont{Deco}}, \bibnamefont{and}
  \bibinfo{author}{\bibfnamefont{D.}~\bibnamefont{Roy}},
  \bibinfo{journal}{Trends in Cognitive Sciences}
  \textbf{\bibinfo{volume}{21}}, \bibinfo{pages}{509 } (\bibinfo{year}{2017}),
  ISSN \bibinfo{issn}{1364-6613}.

\bibitem[{\citenamefont{Nicolaou et~al.}(2017)\citenamefont{Nicolaou, Riecke,
  and Motter}}]{Nicolaou:2017}
\bibinfo{author}{\bibfnamefont{Z.~G.} \bibnamefont{Nicolaou}},
  \bibinfo{author}{\bibfnamefont{H.}~\bibnamefont{Riecke}}, \bibnamefont{and}
  \bibinfo{author}{\bibfnamefont{A.~E.} \bibnamefont{Motter}},
  \bibinfo{journal}{Physical review letters} \textbf{\bibinfo{volume}{119}},
  \bibinfo{pages}{244101} (\bibinfo{year}{2017}).

\bibitem[{\citenamefont{Battogtokh and Kuramoto}(2000)}]{Battogtokh:2000}
\bibinfo{author}{\bibfnamefont{D.}~\bibnamefont{Battogtokh}} \bibnamefont{and}
  \bibinfo{author}{\bibfnamefont{Y.}~\bibnamefont{Kuramoto}},
  \bibinfo{journal}{Physical Review E} \textbf{\bibinfo{volume}{61}},
  \bibinfo{pages}{3227} (\bibinfo{year}{2000}).

\bibitem[{\citenamefont{Schmidt and Krischer}(2015)}]{Schmidt:2015}
\bibinfo{author}{\bibfnamefont{L.}~\bibnamefont{Schmidt}} \bibnamefont{and}
  \bibinfo{author}{\bibfnamefont{K.}~\bibnamefont{Krischer}},
  \bibinfo{journal}{Physical Review Letters} \textbf{\bibinfo{volume}{114}},
  \bibinfo{pages}{034101} (\bibinfo{year}{2015}).

\bibitem[{\citenamefont{Daido and Nakanishi}(2004)}]{Daido:2004}
\bibinfo{author}{\bibfnamefont{H.}~\bibnamefont{Daido}} \bibnamefont{and}
  \bibinfo{author}{\bibfnamefont{K.}~\bibnamefont{Nakanishi}},
  \bibinfo{journal}{Phys. Rev. Lett.} \textbf{\bibinfo{volume}{93}},
  \bibinfo{pages}{104101} (\bibinfo{year}{2004}).

\bibitem[{\citenamefont{Thakur et~al.}(2014)\citenamefont{Thakur, Sharma, and
  Sen}}]{Bhumika:14}
\bibinfo{author}{\bibfnamefont{B.}~\bibnamefont{Thakur}},
  \bibinfo{author}{\bibfnamefont{D.}~\bibnamefont{Sharma}}, \bibnamefont{and}
  \bibinfo{author}{\bibfnamefont{A.}~\bibnamefont{Sen}},
  \bibinfo{journal}{Phys. Rev. E} \textbf{\bibinfo{volume}{90}},
  \bibinfo{pages}{042904} (\bibinfo{year}{2014}).

\bibitem[{\citenamefont{Zou et~al.}(2013)\citenamefont{Zou, Senthilkumar, Zhan,
  and Kurths}}]{Zou:13}
\bibinfo{author}{\bibfnamefont{W.}~\bibnamefont{Zou}},
  \bibinfo{author}{\bibfnamefont{D.~V.} \bibnamefont{Senthilkumar}},
  \bibinfo{author}{\bibfnamefont{M.}~\bibnamefont{Zhan}}, \bibnamefont{and}
  \bibinfo{author}{\bibfnamefont{J.}~\bibnamefont{Kurths}},
  \bibinfo{journal}{Phys. Rev. Lett.} \textbf{\bibinfo{volume}{111}},
  \bibinfo{pages}{014101} (\bibinfo{year}{2013}).

\bibitem[{\citenamefont{Zou et~al.}(2015)\citenamefont{Zou, Senthilkumar,
  Nagao, Kiss, Tang, Koseska, Duan, and Kurths}}]{Zou:15}
\bibinfo{author}{\bibfnamefont{W.}~\bibnamefont{Zou}},
  \bibinfo{author}{\bibfnamefont{D.}~\bibnamefont{Senthilkumar}},
  \bibinfo{author}{\bibfnamefont{R.}~\bibnamefont{Nagao}},
  \bibinfo{author}{\bibfnamefont{I.~Z.} \bibnamefont{Kiss}},
  \bibinfo{author}{\bibfnamefont{Y.}~\bibnamefont{Tang}},
  \bibinfo{author}{\bibfnamefont{A.}~\bibnamefont{Koseska}},
  \bibinfo{author}{\bibfnamefont{J.}~\bibnamefont{Duan}}, \bibnamefont{and}
  \bibinfo{author}{\bibfnamefont{J.}~\bibnamefont{Kurths}},
  \bibinfo{journal}{Nature communications} \textbf{\bibinfo{volume}{6}}
  (\bibinfo{year}{2015}).

\bibitem[{\citenamefont{Chabanol et~al.}(1997)\citenamefont{Chabanol, Hakim,
  and Rappel}}]{Chabanol:97}
\bibinfo{author}{\bibfnamefont{M.-L.} \bibnamefont{Chabanol}},
  \bibinfo{author}{\bibfnamefont{V.}~\bibnamefont{Hakim}}, \bibnamefont{and}
  \bibinfo{author}{\bibfnamefont{W.-J.} \bibnamefont{Rappel}},
  \bibinfo{journal}{Physica D: Nonlinear Phenomena}
  \textbf{\bibinfo{volume}{103}}, \bibinfo{pages}{273} (\bibinfo{year}{1997}).

\bibitem[{\citenamefont{Glendinning and Proctor}(1993)}]{Glendinning:93}
\bibinfo{author}{\bibfnamefont{P.}~\bibnamefont{Glendinning}} \bibnamefont{and}
  \bibinfo{author}{\bibfnamefont{M.}~\bibnamefont{Proctor}},
  \bibinfo{journal}{International Journal of Bifurcation and Chaos}
  \textbf{\bibinfo{volume}{3}}, \bibinfo{pages}{1447} (\bibinfo{year}{1993}).

\end{thebibliography}


\end{document}